\documentclass
[superscriptaddress,secnumarabic,amssymb,amsmath,nobibnotes,aps,prd,showkeys,showpacs,nofootinbib,onecolumn,12pt]{revtex4}%
\usepackage{graphicx}
\usepackage{epsf}
\usepackage{bm}
\usepackage{amsmath}
\usepackage{amsfonts}
\usepackage{amssymb}%
\providecommand{\U}[1]{\protect\rule{.1in}{.1in}}

\newcommand{\be}{\begin{equation}}
\newcommand{\ee}{\end{equation}}

\newcommand{\mincir}{\raise
-3.truept\hbox{\rlap{\hbox{$\sim$}}\raise4.truept\hbox{$<$}\ }}
\newcommand{\magcir}{\raise
-3.truept\hbox{\rlap{\hbox{$\sim$}}\raise4.truept\hbox{$>$}\ }}

\begin{document}
\title{Global dynamics of the hyperbolic Chiral-Phantom model}
\author{Andronikos Paliathanasis}
\email{anpaliat@phys.uoa.gr}
\affiliation{Institute of Systems Science, Durban University of Technology, Durban 4000,
South Africa}
\affiliation{Instituto de Ciencias F\'{\i}sicas y Matem\'{a}ticas, Universidad Austral de
Chile, Valdivia, Chile}
\author{Genly Leon}
\email{genly.leon@ucn.cl}
\affiliation{Departamento de Matem\'{a}ticas, Universidad Cat\'{o}lica del Norte, Avda.
Angamos 0610, Casilla 1280 Antofagasta, Chile}

\begin{abstract}
We perform a detailed analysis of the asymptotic behavior of a multifield cosmological model with phantom terms. Specifically, we consider the Chiral-Phantom model consisting of two scalar fields with a mixed kinetic term, while one scalar field has negative kinetic energy, that is, it has phantom properties. We show that the Hubble function can change sign, and we study the global evolution of the field equation in the finite and infinity regions with the use of Poincar\'{e} variables. We find that in the Chiral-Phantom model the limit of the quintessence scalar field is recovered while the cosmological evolution differs from the standard hyperbolic theory. Finally, the linear cosmological perturbations are studied.

\end{abstract}
\keywords{Multifield Cosmology; Chiral Cosmology; Phantom field; Asympotic behavior;
Dynamical analysis.}
\pacs{98.80.-k, 95.35.+d, 95.36.+x}
\date{\today}
\maketitle

\section{Introduction}

\label{sec1}

Scalar field inflation is the main mechanism that is used to explain the
homogeneity and isotropy of the present universe. The scalar field dominates
in driving the dynamics and explaining the expansion era \cite{Aref1,guth}.
Moreover, the scalar field inflationary models are mainly defined on
homogeneous spacetimes or background spaces with small inhomogeneities
\cite{st1,st2}. Multiscalar field cosmological models because of the
additional degrees of freedom and the different dynamics provide an
alternative approach on the description of the early inflationary era of our
universe. The existence of multifield during the inflationary epoch provides
non-adiabatic field perturbations which can lead to detectable
non-Gaussianities in the power spectrum \cite{mf1,mf2,mf3}. Moreover,
the multifield inflationary model provides a different exit from the inflationary
era. The values of the scalar fields during the beginning and at the exit of
the inflation are not necessarily the same which can lead to a different number of
e-folds and affect the curvature perturbations \cite{mf4,mf5}.

In addition, multiscalar field theories can provide a mechanism for the
description of the late-time acceleration phase of our universe
\cite{dm1,dm2,dm3}. Indeed, the so-called quintom model consists of two scalar
fields where one of the two fields is quintessence while the second scalar
field is phantom \cite{kj1} which means that the energy \cite{qq2,aa2,aa3} can
lead to negative energy density. The main characteristic of this model which
makes it of special interest in cosmological studies is that the effective
parameter for the equation of state for the cosmological fluid can cross the
phantom divide line more than once which can be used to explain some of the
recent cosmological observations, while quintom model can be used as a
mechanism to solve the Hubble tension \cite{rev1}. Another multiscalar field
model of special interest is the two-scalar field model with symmetric potential and hyperbolic field space, also known as the Chiral model
\cite{atr6,atr7} and it is inspired by the $\sigma$-model \cite{sigm0}.

In the Chiral model, the dynamics of the scalar fields is defined in a space of
constant curvature, such that mixed kinetic term provides an effective
interaction between the two fields. In contrast to the quintom model where the
interaction between the two fluids is provided only by the potential terms.
The chiral theory has various applications for the description of the inflationary
epoch \cite{vr91,vr92,cher1} which leads to hyperinflation. Furthermore,
it was found that there can be various applications of the Chiral model in the
late-time universe, because it can describe the background dynamics in the
evolution of the universe, such that to introduce a matter-dominated era,
which means that the Chiral model can be seen as a unified dark model \cite{and1}.
There are various studies in the literature where the field equations are
explicitly solved for different functional forms of the potential function in
Chiral model \cite{ans1,ans2,ans3,ans4,ans5,ans5b}, while some extensions of the chiral model with more scalar fields have been studied before in
\cite{ans6,ans7}. Moreover, in \cite{ans8} an extension of the Chiral model in the
five-dimensional Einstein Gauss-Bonnet theory was considered and new exact
solutions were found.

The effective fluid in the Chiral model has an equation of state parameter
with lower bound the~minus one, in the quantum level there are transitions
such that the parameter for the effective equation of state to cross the
phantom divide line \cite{and2}. Inspired by the latter in \cite{and3},
generalized Chiral models where at least one of the scalar fields has 
negative kinetic energy was proposed. As a primary study of the dynamics of
the background space a Chiral-Phantom model was found with the property that
the equation of state parameter the cosmological fluid crosses twice the
phantom divide line during the evolution without the appearance of ghosts
\cite{and3}. However, when one of the two fields is a phantom, the Hubble
function can change the sign, however, such a case has not been studied before in
\cite{and3}. In this work, we are interested in the global dynamics and the
asymptotic solutions for the Chiral-Phantom model.

The plan of the paper is as follows. In Section \ref{sec2}, we define the
Chiral-Phantom cosmological model of our consideration in a spatially flat
Friedmann--Lema\^{\i}tre--Robertson--Walker \ (FLRW) background space and we
present the gravitational field equations. Furthermore, we define new
dimensionless variables and we write the field equations in an equivalent
system of algebraic-differential equations. Section \ref{sec3} includes the
main results of this analysis where we study the asymptotic behavior for the
cosmological field equations in a different parametrization from that of
$H$-normalization. The asymptotic behavior is investigated in the finite and
infinite regions. Moreover, in Section \ref{per1} we study the linear
cosmological perturbations in the Newtonian gauge. Specifically, we derive the
perturbation equations and we investigate the evolution for the perturbations
of the two coupled scalar fields with the background space described by the
asymptotic solutions. Finally, in Section \ref{con00} we summarize our results
and draw our conclusions.

\section{Chiral-Phantom model}

\label{sec2}

We consider\ an extension of the Chiral cosmological model; specifically, we
assume Einstein's General Relativity with two scalar fields with kinetic terms
defined in the hyperbolic plane, while the second field has a negative kinetic term, that is, the Gravitational Action integral has the following form%
\begin{equation}
S=\int\sqrt{-g}\left(  R-\frac{1}{2}g^{\mu\nu}\nabla_{\mu}\phi\nabla_{\nu}%
\phi+\frac{1}{2}g^{\mu\nu}e^{\kappa\phi}\nabla_{\mu}\psi\nabla_{\nu}%
\psi-V\left(  \phi\right)  \right)  . \label{gd.01}%
\end{equation}
Parameter $\kappa$ is the coupling constant and defines the curvature of the
two-dimensional hyperbolic space in which the motion of the scalar field
occurs. In the limit $\kappa=0$, this two-dimensional space becomes flat and
the gravitational action integral (\ref{gd.01}) is that of the quintom model. In
our analysis, we focus on the case of~$\sigma-$model where $\kappa\neq0$.

For the cosmological background space of a spatially flat FLRW spacetime%
\begin{equation}
ds^{2}=-dt^{2}+a\left(  t\right)  ^{2}\left(  dx_{1}^{2}+dx_{2}^{2}+dx_{2}%
^{2}\right)  , \label{ss.01}%
\end{equation}
where $a\left(  t\right)  $ is the scale factor which is the radius of the
three-dimensional hypersurface, variation with respect to the metric tensor in
(\ref{gd.01}) provides the field equations which are \cite{and3}
\begin{align}
3H^{2}  &  =\frac{1}{2}\dot{\phi}^{2}-\frac{1}{2}e^{\kappa\phi}\dot{\psi}%
^{2}+V\left(  \phi\right)  ,\label{ss.05}\\
-\left(  2\dot{H}+3H^{2}\right)   &  =\frac{1}{2}\dot{\phi}^{2}-\frac{1}%
{2}e^{\kappa\phi}\dot{\psi}^{2}-V\left(  \phi\right)  . \label{ss.06}%
\end{align}
where dot means derivative with respect to the time variable and $H\left(
t\right)  =\frac{\dot{a}}{a}$ is the Hubble function. Furthermore, variation
with respect to the scalar fields leads to the Klein-Gordon equations%
\begin{equation}
\left(  \ddot{\phi}+3H\dot{\phi}\right)  +\frac{1}{2}\kappa e^{\kappa\phi}%
\dot{\psi}^{2}+V_{,\phi}=0~, \label{ss.07}%
\end{equation}%
\begin{equation}
\ddot{\psi}+3H\dot{\psi}+\kappa\dot{\phi}\dot{\psi}=0~, \label{ss.08}%
\end{equation}
where we have assumed that the scalar fields inherit the symmetries of the
spacetime, that is, $\phi\left(  x^{\mu}\right)  =\phi\left(  t\right)  $ and
$\psi\left(  x^{\mu}\right)  =\phi\left(  t\right)  $. Equation (\ref{ss.08})
becomes $\left(  a^{3}e^{\kappa\phi}\dot{\psi}\right)  ^{\cdot}=0$, which
provides the conservation law $I_{0}=a^{3}e^{\kappa\phi}\dot{\psi}$.

\subsection{Dimensionless variables}

From equation (\ref{ss.05}) we observe that $H\left(  t\right)  $ can take the
value zero, i.e. $H\left(  t\right)  =0$, which means that we can not continue
with the standard $H-$normalization \cite{cop1} for the definition of the
dimensionless variables. Thus, we introduce a new parametrization and we
define the new variables \cite{gg1}%
\begin{equation}
x=\frac{\dot{\phi}}{\sqrt{6\left(  1+H^{2}\right)  }}~,~y^{2}=\frac{V\left(
\phi\right)  }{3\left(  1+H^{2}\right)  }~,~z=e^{\frac{\kappa}{2}\phi}%
\frac{\dot{\psi}}{\sqrt{6\left(  1+H^{2}\right)  }}~,~\eta^{2}=\frac{H^{2}%
}{1+H^{2}}. \label{ss.09}%
\end{equation}

In these new variables, the constraint equation (\ref{ss.05}) reads
\begin{equation}
\eta^{2}-x^{2}-y^{2}+z=0, \label{ss.10}%
\end{equation}
while the evolution equations are
\begin{align}
x^{\prime}  &  =\frac{1}{2}\left(  3x^{3}\eta-3x\eta\left(  2+y^{2}+z-\eta
^{2}\right)  -\sqrt{6}\left(  z\kappa+\lambda y^{2}\right)  \right)
,\label{ss.11}\\
y^{\prime}  &  =\frac{1}{2}y\left(  3\eta^{3}-3\eta\left(  y^{2}%
+z-x^{2}\right)  +\sqrt{6}\lambda x\right)  ,\label{ss.12}\\
z^{\prime}  &  =z\left(  3\eta^{3}-3\left(  2-x^{2}+y^{2}+z\right)  \eta
-\sqrt{6}\kappa x\right)  ,\label{ss.13}\\
\eta^{\prime}  &  =\frac{3}{2}\left(  \eta^{2}-1\right)  \left(  x^{2}%
-y^{2}-z+\eta^{2}\right)  , \label{ss.14}%
\end{align}
where a prime notes derivative with respect to the new independent variable
$\tau,~d\tau=\sqrt{1+H^{2}}dt$. Function $\lambda$ is defined as
$\lambda=\left(  \ln V\left(  \phi\right)  \right)  _{,\phi}$ with the
evolution equation%
\begin{equation}
\lambda^{\prime}=\sqrt{6}\lambda^{2}x\left(  \Gamma\left(  \lambda\right)
-1\right)  ,\Gamma\left(  \lambda\right)  =\frac{V_{,\phi\phi}V}{\left(
V_{,\phi}\right)  ^{2}}\text{.} \label{ss.15}%
\end{equation}
For the scalar field potential we select the exponential $V\left(
\phi\right)  =V_{0}e^{\lambda_{0}\phi}$ which gives $\lambda^{\prime}=0$, that
is, $\lambda$ is a constant, i.e. $\lambda=\lambda_{0}$.

From (\ref{ss.10}) it follows that the variables are not bounded and they can
take values in the range of real numbers, except the variable $y$ which
we assume that is positive, i.e. $y>0$.

\section{Global dynamics}

\label{sec3}

This work focus on the study of the global dynamics for the Chiral-Phantom
model. Specifically, we investigate the asymptotic behavior for the
dimensionless algebraic-differential system (\ref{ss.10})-(\ref{ss.14}). Any
stationary point of the latter system describes an asymptotic exact solution
for the cosmological field equations with the effective equation of state
parameter $w_{eff}\left(  x,y,z,\eta\right)  =1-\frac{2y^{2}}{x^{2}+y^{2}-z}$.
Previous studies related to analyzes of the dynamics of multi-field models can
be found for instance in \cite{mmf1,mmf2,mmf3,mmf4}, while for the Chiral
model it can be found \cite{and1}.

\subsection{Local analysis}

From the constraint equation (\ref{ss.10}) we find $z=x^{2}+y^{2}-\eta^{2}$,
which can be used to reduce the dynamical system in the following dynamical
system of three first-order differential equations%
\begin{align}
x^{\prime}  &  =\frac{1}{2}\left(  3x^{3}\eta-6x\eta\left(  1+y^{2}\right)
+\sqrt{6}\kappa\eta^{2}-\sqrt{6}\left(  x^{2}\kappa+y^{2}\left(
\kappa+\lambda\right)  \right)  \right)  ,\label{ss.16}\\
y^{\prime}  &  =3y\eta\left(  \eta^{2}-y^{2}\right)  +\frac{\sqrt{6}}%
{2}\lambda xy,\label{ss.17}\\
\eta^{\prime}  &  =\frac{3}{2}\left(  \eta^{2}-1\right)  \left(  \eta
^{2}-y^{2}\right)  . \label{ss.18}%
\end{align}

We shall find \ all the points $P=P\left(  x\left(  P\right)  ,y\left(
P\right)  ,\eta\left(  P\right)  \right)  $ of the phase space in which the
right hand side of the later system are zero. Points $P$ are stationary points
for the dynamical system. Hence, from the algebraic equations:%
\begin{align}
\left(  3x^{3}\eta-6x\eta\left(  1+y^{2}\right)  +\sqrt{6}\kappa\eta^{2}%
-\sqrt{6}\left(  x^{2}\kappa+y^{2}\left(  \kappa+\lambda\right)  \right)
\right)   &  =0,\label{ss.19}\\
3y\eta\left(  \eta^{2}-y^{2}\right)  +\frac{\sqrt{6}}{2}\lambda xy  &
=0,\label{ss.20}\\
\frac{3}{2}\left(  \eta^{2}-1\right)  \left(  \eta^{2}-y^{2}\right)   &  =0,
\label{ss.21}%
\end{align}
we obtain the following singular points:

$P_{1}=\left(  1,0,1\right)  $ describes a universe dominated by the kinetic
term of the scalar field $\phi$, while the second field $\psi$ and the scalar
field potential $V\left(  \phi\right)  $ do not contribute in the cosmological
fluid. The asymptotic solution is that of the stiff fluid source,
$w_{eff}\left(  P_{1}\right)  =1$. The eigenvalues of the linearized system
are $e_{1}\left(  P_{1}\right)  =6~,~e_{2}\left(  P_{1}\right)  =-\sqrt
{6}\kappa,~e_{3}\left(  P_{1}\right)  =\frac{1}{2}\left(  6+\sqrt{6}%
\lambda\right)  $. Therefore, the point is a source when $\kappa<0$%
,~$\lambda>-\sqrt{6}$ or a saddle point when $\kappa>0$ and $\lambda\,$ \ arbitrary.

$P_{2}=\left(  -1,0,1\right)  $ has the same physical properties as point
$P_{1}$. However, the eigenvalues of the linearized system are~$e_{1}\left(
P_{2}\right)  =6~,~e_{2}\left(  P_{2}\right)  =\sqrt{6}\kappa,~e_{3}\left(
P_{2}\right)  =\frac{1}{2}\left(  6-\sqrt{6}\lambda\right)  $ from where we
infer that the point is a source for $\kappa>0,~\lambda<\sqrt{6}$, while for
$\kappa<0$,~$P_{2}$ is a saddle point.

$P_{3}=\left(  1,0,-1\right)  $ with physical properties similar to $P_{1}$
and eigenvalues $e_{1}\left(  P_{3}\right)  =-6~,~e_{2}\left(  P_{3}\right)
=-\sqrt{6}\kappa,~e_{3}\left(  P_{3}\right)  =\frac{1}{2}\left(  -6+\sqrt
{6}\lambda\right)  $. Hence, $P_{3}$ is an attractor, for $\kappa>0$%
,~$\lambda<\sqrt{6}$ otherwise is a saddle point.

$P_{4}=\left(  -1,0,-1\right)  $ with physical solution similar to $P_{1}$ and
eigenvalues $e_{1}\left(  P_{4}\right)  =-6~,~e_{2}\left(  P_{4}\right)
=\sqrt{6}\kappa,~e_{3}\left(  P_{4}\right)  =\frac{1}{2}\left(  -6-\sqrt
{6}\lambda\right)  $. Hence, $P_{4}$ is an attractor, for $\kappa>0$%
,~$\lambda>\sqrt{6}$. Otherwise, it is a saddle point.

\begin{figure}[ptb]
\centering\includegraphics[width=0.5\textwidth]{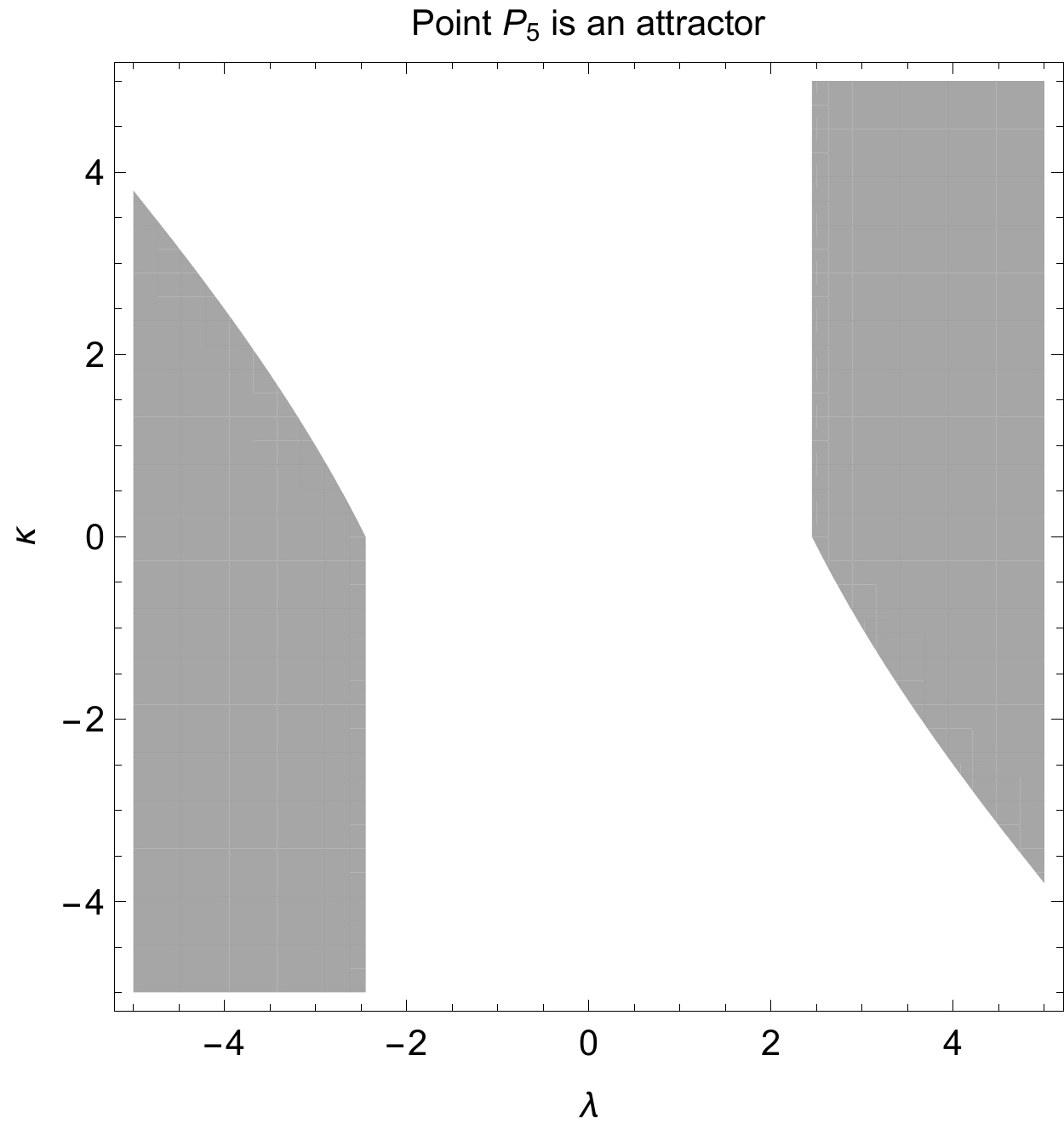} \caption{Region space
of the free parameters $\left\{  \lambda,\kappa\right\}  $ in which point
$P_{5}$ is an attractor.}%
\label{pr1}%
\end{figure}

$P_{5}=\left(  \frac{\lambda}{\sqrt{6}},\sqrt{1-\frac{\lambda^{2}}{6}%
},-1\right)  $ describes a quintessence model where the second scalar field
$\psi$ does not contribute in the cosmological fluid. The equation of state
parameter is $w_{eff}=\frac{\lambda^{2}}{3}-1$ which means that it describes
an inflationary universe for $-\sqrt{2}<\lambda<\sqrt{2}$. The eigenvalues of
the linearized system are $e_{1}\left(  P_{5}\right)  =-\lambda^{2}%
~,~e_{2}\left(  P_{5}\right)  =\frac{1}{2}\left(  6-\lambda^{2}\right)
,~e_{3}\left(  P_{5}\right)  =\left(  6-\kappa\lambda-\lambda^{2}\right)  $.
Therefore, the stationary point is an attractor when $\left\{  \lambda
<-\sqrt{6}~,~\lambda>-\frac{\kappa}{2}+\frac{\sqrt{24+\kappa^{2}}}{2}%
,~\kappa<0\right\}  ~$or $\left\{  \lambda>\sqrt{6}~,~\lambda<-\frac{\kappa
}{2}-\frac{\sqrt{24+\kappa^{2}}}{2},~\kappa>0\right\}  $. Otherwise, $P_{5}$
is a saddle point. The region space of the variables $\left\{  \lambda
,\kappa\right\}  $ in which point $P_{5}$ is an attractor is presented in Fig.
\ref{pr1}. 

$P_{6}=\left(  -\frac{\lambda}{\sqrt{6}},\sqrt{1-\frac{\lambda^{2}}{6}%
},1\right)  $ with physical properties similar to $P_{5}$. The eigenvalues of
the linearized system are $e_{1}\left(  P_{6}\right)  =\lambda^{2}%
~,~e_{2}\left(  P_{6}\right)  =\frac{1}{2}\left(  \lambda^{2}-6\right)
,~e_{3}\left(  P_{6}\right)  =-\left(  6-\kappa\lambda-\lambda^{2}\right)  $
from where we infer that $P_{6}$ can not be a stable point while it is a
saddle point for $\lambda^{2}<6$ or $\left\{  \lambda<0,\kappa<\frac
{6-\lambda^{2}}{\lambda}\right\}  $ or $\left\{  \lambda>0,\kappa
>\frac{6-\lambda^{2}}{\lambda}\right\}  $.

\begin{figure}[ptb]
\centering\includegraphics[width=0.5\textwidth]{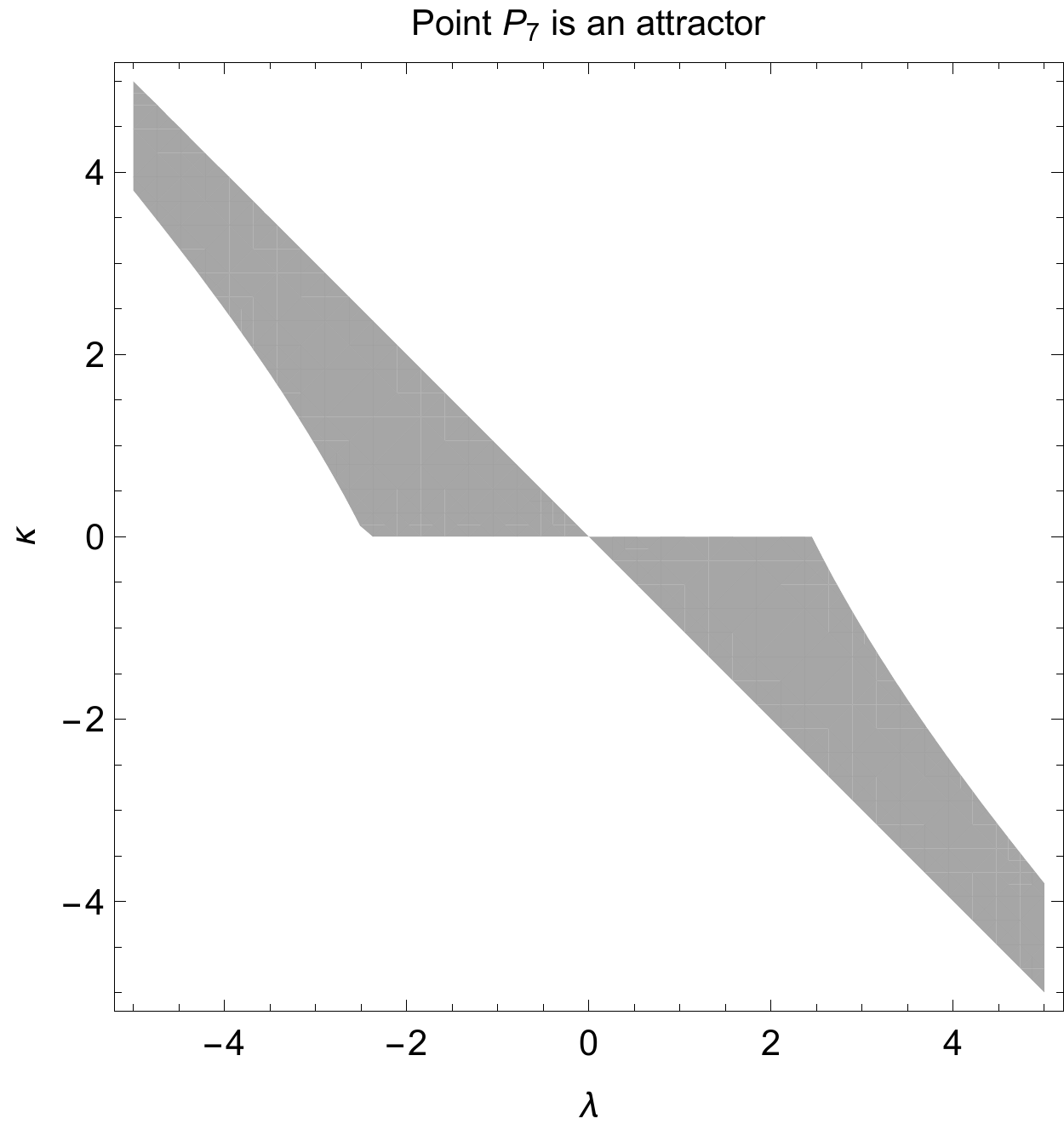} \caption{Region space
of the free parameters $\left\{  \lambda,\kappa\right\}  $ in which point
$P_{7}$ is an attractor.}%
\label{pr2}%
\end{figure}

$P_{7}=\left(  \frac{\sqrt{6}}{\kappa+\lambda},\sqrt{\frac{\kappa}%
{\kappa+\lambda}},-1\right)  $ \ describes the so-called hyperinflation exact
solution in which both the scalar fields contribute to the cosmological fluid.
The equation of state parameter is $w_{eff}\left(  P_{7}\right)
=1-\frac{2\kappa}{\kappa+\lambda}$ which means $w_{eff}\left(  P_{7}\right)
<-\frac{1}{3}$ for $\ \left\{  \lambda<0~|~\kappa>-\lambda\text{ or }%
\kappa<2\lambda\right\}  $ or~$\left\{  \lambda>0~|~\kappa<-\lambda\text{ or
}\kappa>2\lambda\right\}  $. The asymptotic solution describes an inflationary
universe. Moreover, it is important to mention that the point is physically
acceptable when $\frac{\kappa}{\kappa+\lambda}>0~$and exist when
$\kappa+\lambda\neq0$. The eigenvalues of the linearized system are
$e_{1}\left(  P_{7}\right)  =-\frac{6\lambda}{\kappa+\lambda}~,~e_{2}\left(
P_{7}\right)  =\frac{3\kappa}{\kappa+\lambda}+\frac{\sqrt{3\kappa}}%
{\kappa+\lambda}\left(  27\kappa-4\left(  \kappa^{2}-6\right)  \lambda
-8\kappa\lambda^{2}-4\lambda^{3}\right)  ^{\frac{1}{2}}$, ~$e_{3}\left(
P_{7}\right)  =\frac{3\kappa}{\kappa+\lambda}-\frac{\sqrt{3\kappa}}%
{\kappa+\lambda}\left(  27\kappa-4\left(  \kappa^{2}-6\right)  \lambda
-8\kappa\lambda^{2}-4\lambda^{3}\right)  ^{\frac{1}{2}}$. The region space of
the variables $\left\{  \lambda,\kappa\right\}  $ in which point $P_{7}$ is an
attractor is presented in Fig. \ref{pr2}. 

$P_{8}=\left(  -\frac{\sqrt{6}}{\kappa+\lambda},\sqrt{\frac{\kappa}%
{\kappa+\lambda}},1\right)  $ has physical properties similar to point $P_{7}%
$. The eigenvalues of the linearized system are $e_{1}\left(  P_{8}\right)
=\frac{6\lambda}{\kappa+\lambda}~,~e_{2}\left(  P_{7}\right)  =-\frac{3\kappa
}{\kappa+\lambda}+\frac{\sqrt{3\kappa}}{\kappa+\lambda}\left(  27\kappa
-4\left(  \kappa^{2}-6\right)  \lambda-8\kappa\lambda^{2}-4\lambda^{3}\right)
^{\frac{1}{2}}$, ~$e_{3}\left(  P_{7}\right)  =-\frac{3\kappa}{\kappa+\lambda
}-\frac{\sqrt{3\kappa}}{\kappa+\lambda}\left(  27\kappa-4\left(  \kappa
^{2}-6\right)  \lambda-8\kappa\lambda^{2}-4\lambda^{3}\right)  ^{\frac{1}{2}}%
$. Then, it follows that $P_{7}$ is a saddle point.

$P_{9}=\left(  0,0,0\right)  \,\ \ $describes the Minkowski spacetime, where
$H=0$. This point describes the transition where the\ Hubble function changes
the sign. The eigenvalues of the linearized system are found to be zero
$e_{1}\left(  P_{9}\right)  =0~,~e_{2}\left(  P_{9}\right)  =0~$\ and
$e_{3}\left(  P_{9}\right)  =0$. By numerical inspection it is shown this
point is always a saddle point.

\begin{figure}[ptb]
\centering\includegraphics[width=1\textwidth]{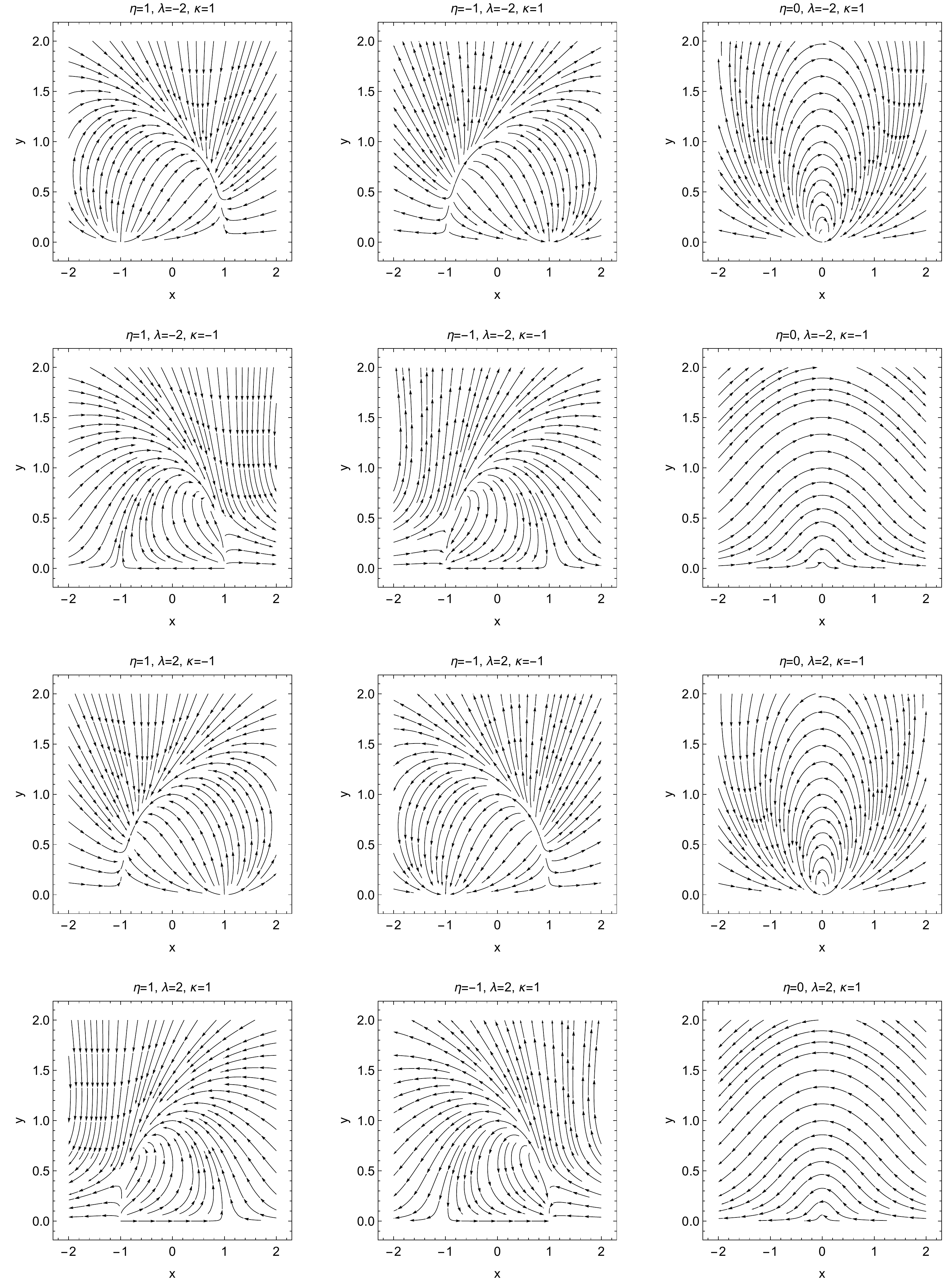}
\caption{Two-dimensional phase space portrait for the dynamical system
(\ref{ss.16})-(\ref{ss.18}) in the $x-y$ plane for $\eta^{2}=1$ and $\eta=0$.}%
\label{pr3}%
\end{figure}

\begin{figure}[ptb]
\centering\includegraphics[width=0.7\textwidth]{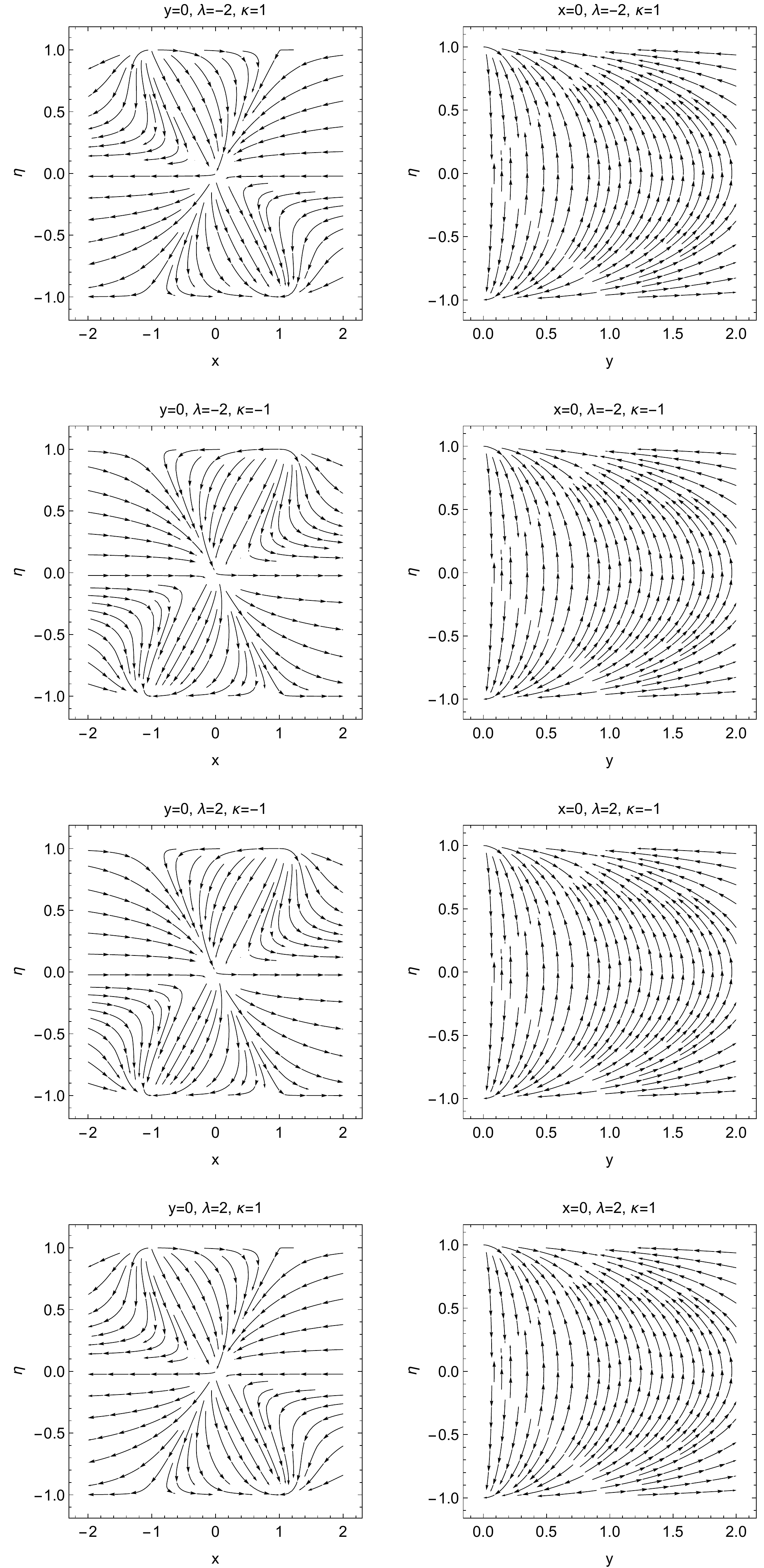}
\caption{Two-dimensional phase space portrait for the dynamical system
(\ref{ss.16})-(\ref{ss.18}) in the $x-\eta$ and $y-\eta$ planes.}%
\label{pr4}%
\end{figure}

\begin{figure}[ptb]
\centering\includegraphics[width=1\textwidth]{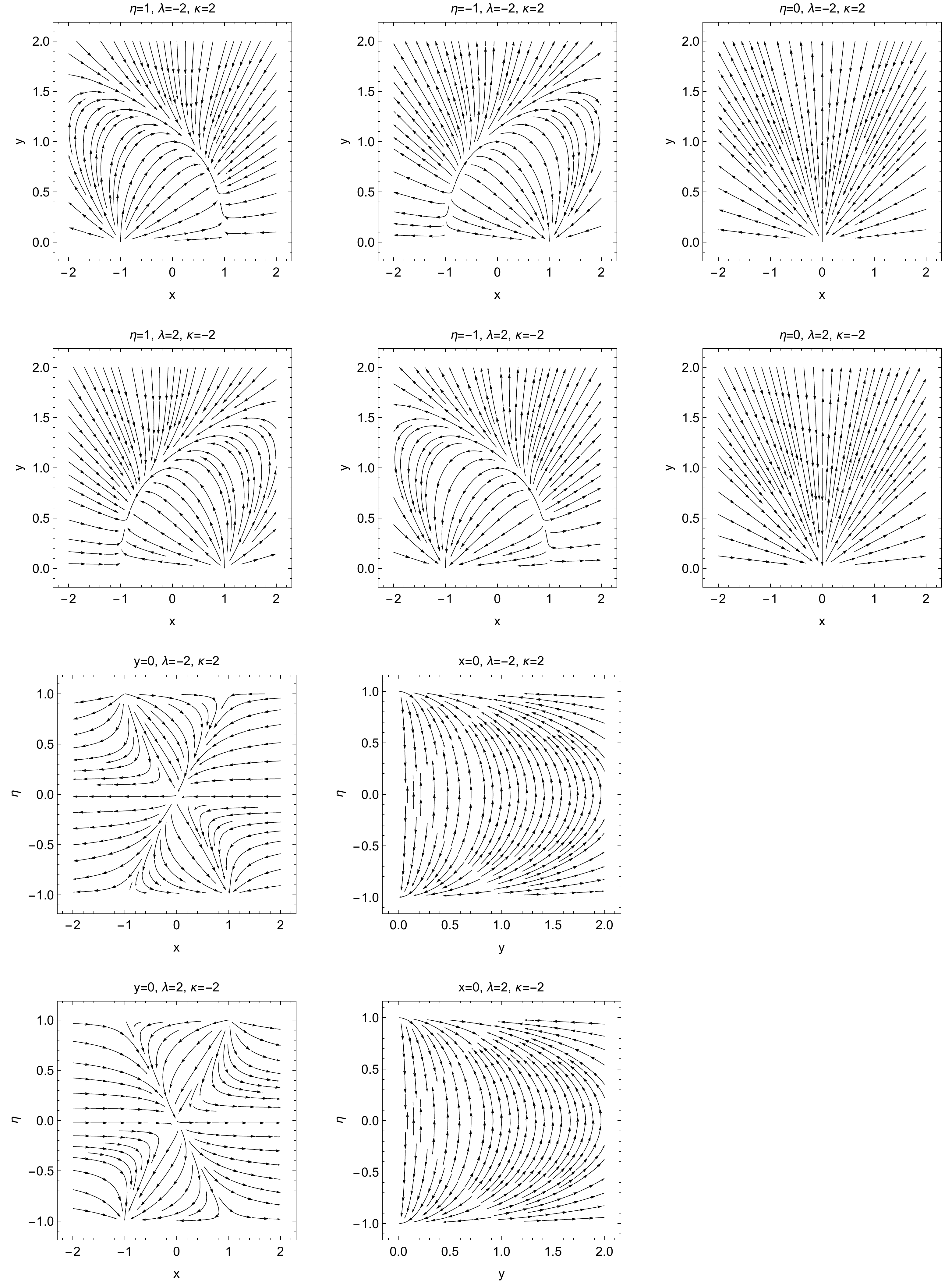}
\caption{Two-dimensional phase space portrait for the dynamical system
(\ref{ss.16})-(\ref{ss.18}) in the $x-y$ plane for $\eta^{2}=1,~\eta=0$, and
in the $x-\eta$ and $y-\eta$ planes, for $\kappa+\lambda=0$. }%
\label{pr7}%
\end{figure}

\begin{figure}[ptb]
\centering\includegraphics[width=0.9\textwidth]{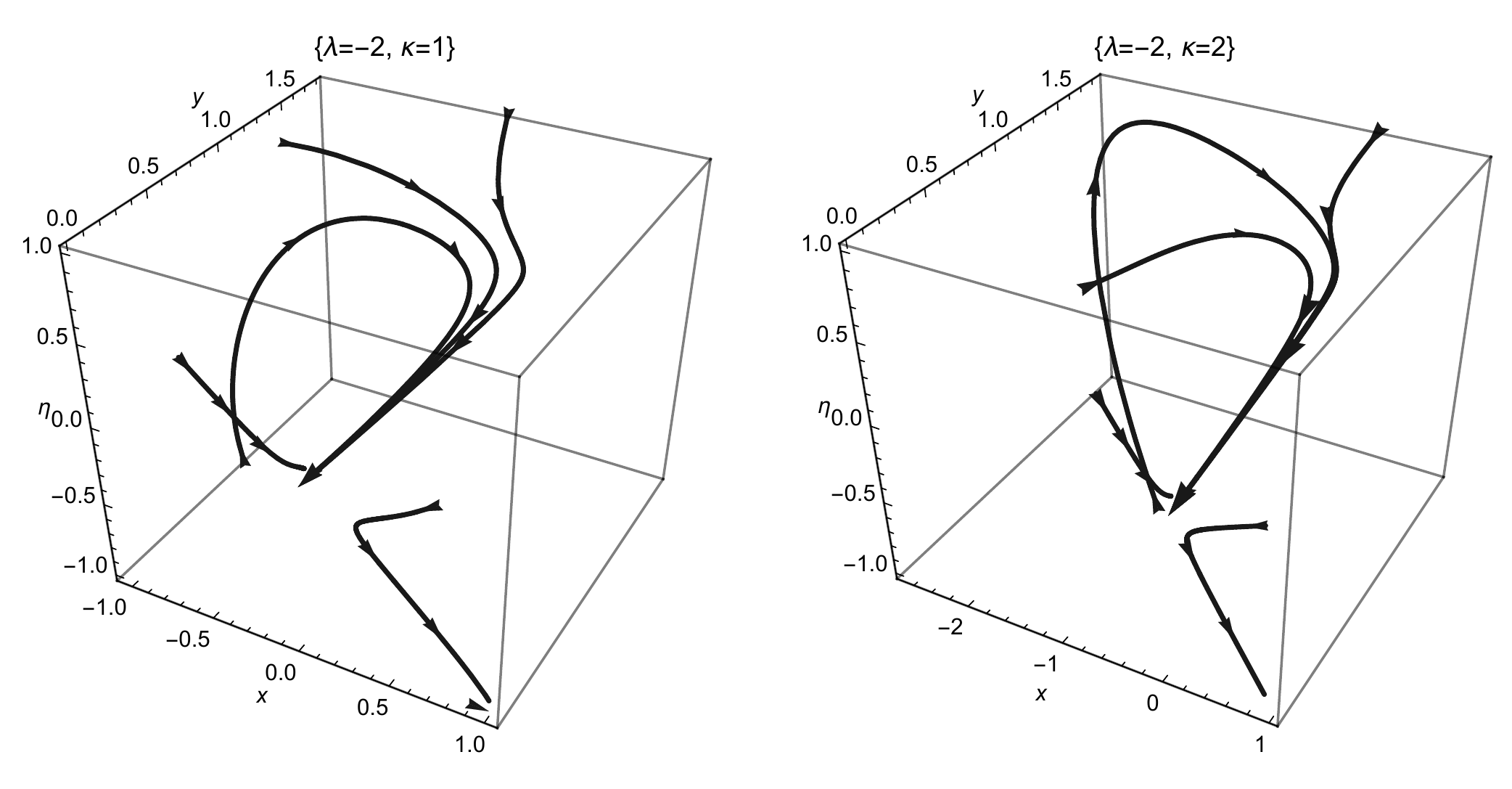} \caption{Trajectories
in the three-dimensional phase space for the dynamical system (\ref{ss.16}%
)-(\ref{ss.18}) for different values of the free parameters $\lambda$ and
$\kappa$.}%
\label{pr6}%
\end{figure}

Figs. \ref{pr3} and \ref{pr4} show two-dimensional phase space portraits for
the dynamical system (\ref{ss.16})-(\ref{ss.18}) for different values of the
free parameters $\lambda,~\kappa$. In Figs. \ref{pr3}, orbits in the $x-y$
plane are displayed; in Fig. \ref{pr4} orbits in the $x-\eta$ and $y-\eta$
planes are presented. In Figs. \ref{pr7} orbits in the special case
$\kappa+\lambda=0$ in a two-dimensional phase space are presented. Moreover,
Fig. \ref{pr6} shows orbits of the three-dimensional phase space of the
dynamical system under study.

\subsection{Analysis at infinity}

We continue our analysis by studying the existence of stationary points at the
infinity. For this\ reason, we perform the change of variables by assuming the
Poincar\'{e} variables%
\begin{equation}
x=\frac{\rho}{\sqrt{1-\rho^{2}}}\cos\chi\cos\theta,~~y=\frac{\rho}%
{\sqrt{1-\rho^{2}}}\sin\chi\cos\theta~,~\eta=~\frac{\rho}{\sqrt{1-\rho^{2}}%
}\sin\theta, \label{ss.23}%
\end{equation}
where $\chi\in\lbrack0,\pi],~\theta\in\lbrack-\frac{\pi}{2},\frac{\pi}{2}]$
and $0\leq\rho\leq1$. Therefore, the dynamical system (\ref{ss.16}%
)-(\ref{ss.18}) reads
\begin{align}
\frac{4}{\rho}\frac{d\rho}{d\sigma}  &  =\sqrt{6}\kappa\left(  \rho
^{2}-1\right)  \left(  \cos\theta+\cos\left(  3\theta\right)  \right)
\cos\chi+3\left(  3\rho^{2}-2-\left(  5\rho^{2}-2\right)  \cos2\theta\right)
\sin\theta\nonumber\\
&  +6\left(  3\rho^{2}-2\right)  \cos^{2}\theta\cos\left(  2\chi\right)
\sin\theta~,~\label{ss.25a}\\
\frac{d\chi}{d\sigma}  &  =\frac{1}{4}\left(  \sqrt{6}\left(  \lambda+\left(
2\kappa+\lambda\right)  \cos\left(  2\theta\right)  \right)  \frac{\sin\chi
}{\cos\theta}+6\sin\theta\sin\left(  2\chi\right)  \right)  ~,\label{ss.26}\\
\frac{d\theta}{d\sigma}  &  =\frac{1}{2}\cos\left(  2\theta\right)  \left(
\sqrt{6}\kappa\cos\chi\sin\theta+6\sin^{2}\chi\cos\theta\right)  ~,
\label{ss.27}%
\end{align}
where the new independent variable $\sigma$ is defined as $d\sigma=\frac
{\sqrt{1-\rho^{2}}}{\rho}d\tau$.

We work at the infinity in which $\rho\rightarrow1^{+}$. As $\rho
\rightarrow1^{+}$ the leading terms in \eqref{ss.25a} are
\begin{equation}
\frac{d\rho}{d \sigma}=\frac{3}{4} \sin(\theta) \left(  2 \cos^{2}(\theta)
\cos(2 \chi)-3 \cos(2 \theta)+1\right)  . \label{ss.radial}%
\end{equation}
Then, we analyze the stability at infinity using the asymptotic equation
\eqref{ss.radial} together with equations \eqref{ss.26} and \eqref{ss.27}. The
radial equation \eqref{ss.radial} does not contain the radial coordinate, so
the singular points can be obtained using the angular equations only. Setting
$\chi^{\prime}(\sigma)= 0$, and $\theta^{\prime}(\sigma)=0$, we obtain the
singular points which are listed below. The stability of these points is
studied by analyzing first the stability of the angular coordinates and then
deducing, from the sign of equation \eqref{ss.radial}, the stability on the
radial direction. Note that from $\frac{d\rho}{d\sigma}=c$, if $c>0$ the
$\rho$ increases and reaches the value one, while when $c<0$, $\rho$ decreases
and goes far from the value $\rho=1$.

\ The stationary points of the two dimensional dynamical system (\ref{ss.26}),
(\ref{ss.27}) in the range $\chi\in\lbrack0,\pi],~\theta\in\lbrack-\frac{\pi
}{2},\frac{\pi}{2}]$ are:

$Q_{1}^{\pm}=\left(  0,\pm\frac{\pi}{4}\right)  $ that describe stiff fluid
solutions, $w_{eff}\left(  Q_{1}^{\pm}\right)  =0$, with $\eta\neq0$. The
eigenvalues of the linearized system are $e_{1}\left(  Q_{1}^{\pm}\right)
=-\sqrt{3}\kappa~,~e_{2}\left(  Q_{2}^{\pm}\right)  =\frac{1}{2}\left(
\sqrt{3}\lambda\pm3\sqrt{2}\right)  $, while from \eqref{ss.radial} we
find,~$\frac{d\rho}{d\sigma}=\pm\frac{3}{2\sqrt{2}}$ as $\rho\rightarrow1^{+}%
$, where we find point $Q_{1}^{+}$ is a sink when $\left\{  \kappa
>0,\lambda<-\sqrt{6}\right\}  $ while $Q_{2}^{-}$ is always a source.

$Q_{2}=\left(  0,0\right)  $ is a stationary point where $\eta=0$, which means
that $H=0$; hence, it changes the sign. The eigenvalues of the linearized
system are $e_{1}\left(  Q_{2}\right)  =\sqrt{\frac{3}{2}}\kappa
~,~e_{2}\left(  Q_{2}\right)  =\sqrt{\frac{3}{2}}\left(  \kappa+\lambda
\right)  $, while from \eqref{ss.radial} we have $\frac{d\rho}{d\sigma}=0$.
Therefore, $e_{1}\left(  Q_{2}\right)  <0,~e_{2}\left(  Q_{2}\right)  <0$ when
$\kappa<0$ and $\lambda<-\kappa$.

$Q_{3}^{\pm}=\left(  \pi,\pm\frac{\pi}{4}\right)  $ have physical properties
similar to $Q_{1}^{\pm}$. The eigenvalues of the linearized system are
$e_{1}\left(  Q_{3}^{\pm}\right)  =\sqrt{3}\kappa~,~e_{2}\left(  Q_{3}^{\pm
}\right)  =\frac{1}{2}\left(  -\sqrt{3}\lambda\pm3\sqrt{2}\right)  $. From
equation \eqref{ss.radial} it follows $\frac{d\rho}{d\sigma}=\pm\frac
{3}{2\sqrt{2}}$ from where we conclude that $Q_{3}^{+}$ is a sink for
$\kappa<0$ and $\lambda>\sqrt{6}$ while $Q_{3}^{-}$ is always a source.

$Q_{4}=\left(  \pi,0\right)  ~$is a point in which $\eta=0$. The eigenvalues
of the linearized system are $e_{1}\left(  Q_{2}\right)  =-\sqrt{\frac{3}{2}%
}\kappa~,~e_{2}\left(  Q_{2}\right)  =-\sqrt{\frac{3}{2}}\left(
\kappa+\lambda\right)  $. Hence, they are negative when $\kappa>0,~\lambda
>-\kappa$. Moreover, from \eqref{ss.radial} it follows $\frac{d\rho}{d\sigma
}=0$.

$Q_{5}^{\pm}=\left(  \pm\arccos\left(  \frac{\lambda}{\sqrt{6}}\right)
,-\frac{\pi}{4}\right)  $ describes a scaling solution with $w_{ff}\left(
Q_{5}^{\pm}\right)  =\frac{\lambda^{2}}{3}-1$, they are the analog of point
$P_{5}$ at infinity. Equation \eqref{ss.radial} gives $\frac{d\rho}{d\sigma
}=-\frac{\lambda^{2}}{4\sqrt{2}}$, while the eigenvalues of the linearized
system are $e_{1}\left(  Q_{5}^{\pm}\right)  =\frac{1}{\sqrt{2}}\left(
6-\lambda\left(  \kappa+\lambda\right)  \right)  ,~e_{2}\left(  Q_{5}^{\pm
}\right)  =\frac{1}{2\sqrt{2}}\left(  6-\lambda^{2}\right)  $. Therefore, we
conclude that for $\left\{  \lambda^{2}<0~,~\kappa\lambda<6-\lambda
^{2}\right\}  $ the points are sources. Otherwise they are saddle points.

$Q_{6}^{\pm}=\left(  \pm\arccos\left(  -\frac{\lambda}{\sqrt{6}}\right)
,\frac{\pi}{4}\right)  $~~are the equivalent points of $P_{6}$ at infinity
where we find $\frac{d\rho}{d\sigma}=\frac{\lambda^{2}}{4\sqrt{2}}$%
,~$e_{1}\left(  Q_{6}^{\pm}\right)  =\frac{3}{2\sqrt{2}}\left(  \lambda
^{2}-2\right)  ,e_{2}\left(  Q_{6}^{\pm}\right)  =\frac{\lambda^{2}%
-\kappa\lambda-6}{\sqrt{2}}$. Therefore, we infer that the points are sinks
when $\lambda^{2}<2$ and $\kappa\lambda>\lambda^{2}-6$.

\begin{figure}[ptb]
\centering\includegraphics[width=0.9\textwidth]{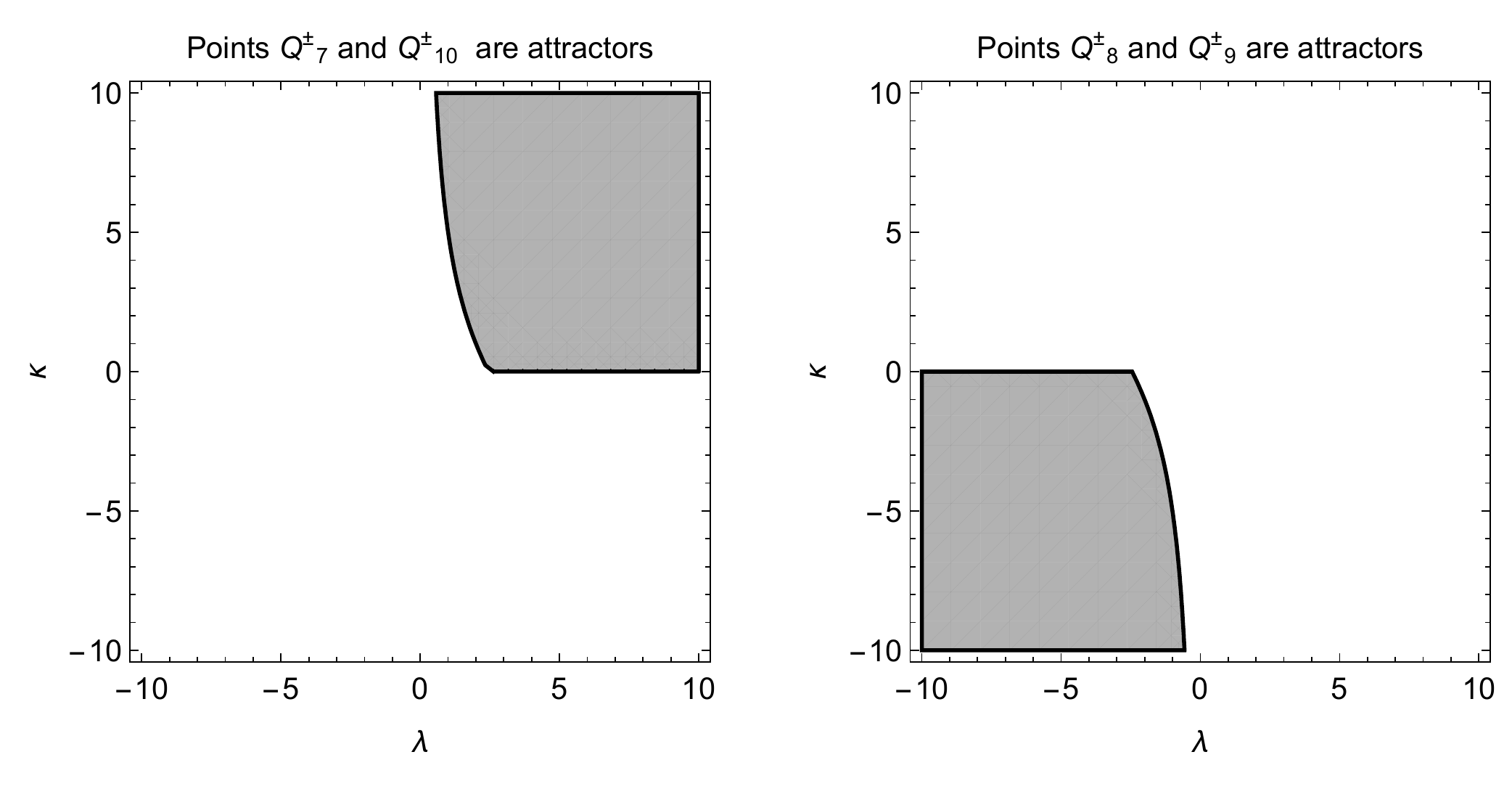} \caption{Region plot
in the space of the free variables~$\left(  \lambda,\kappa\right)  $ where the
stationary points \thinspace$Q_{7}^{\pm},~$\thinspace$Q_{8}^{\pm}$%
,~\thinspace$Q_{8}^{\pm}\ $and \thinspace$Q_{10}^{\pm}$ are attractors. }%
\label{m8p}%
\end{figure}

Furthermore, we find the stationary points \thinspace$Q_{7}^{\pm}=\left(
\pm2\arctan\left(  \delta^{-}\right)  ,2\arctan\left(  \upsilon^{+}\right)
\right)  ,Q_{8}^{\pm}=\left(  \pm2\arctan\left(  \delta^{+}\right)
,-2\arctan\left(  \upsilon^{+}\right)  \right)  $ , $Q_{9}^{\pm}=\left(
\pm2\arctan\left(  \delta^{-}\right)  ,2\arctan\left(  \upsilon^{-}\right)
\right)  $ and $Q_{10}^{\pm}=\left(  \pm2\arctan\left(  \delta^{+}\right)
,-2\arctan\left(  \upsilon^{-}\right)  \right)  $, where $\delta^{\pm}%
=\sqrt{\frac{12+\kappa\left(  \kappa+\lambda\right)  \pm2\sqrt{6\left(
6+\kappa\left(  \kappa+\lambda\right)  \right)  }}{\left(  \kappa
+\lambda\right)  }}~~,~\upsilon^{\pm}=\frac{\sqrt{6+\kappa\left(
\kappa+\lambda\right)  }\pm\sqrt{6+\left(  \kappa+\lambda\right)  \left(
2\kappa+\lambda\right)  }}{\left(  \kappa+\lambda\right)  }$. The points are
real when $\left\{  \kappa<0,~\lambda>0\text{ and }\kappa+\lambda>0\right\}  $
or~$\left\{  \kappa>0,~\lambda<0\text{ and }\kappa+\lambda<0\right\}  $. The
asymptotic solutions describe scaling solutions, with the same physics of
point $P_{7}$. In Fig. \ref{m8p} the regions of $\left(  \lambda
,\kappa\right)  $ in which the points are sinks are present.

\begin{figure}[ptb]
\centering\includegraphics[width=0.9\textwidth]{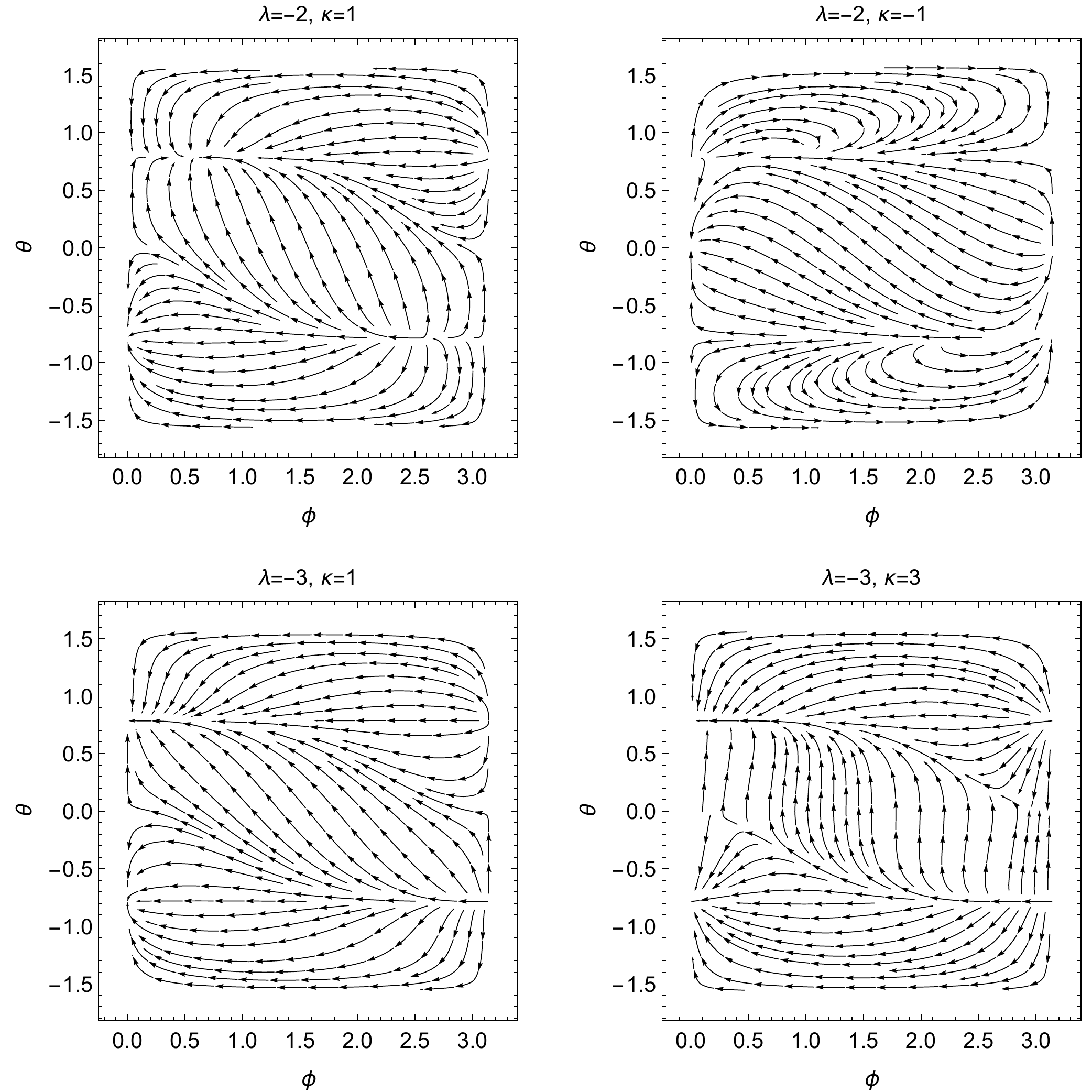} \caption{Phase-space
portraits for the dynamical system (\ref{ss.26}), (\ref{ss.27}) for different
values of the free parameters $\left(  \lambda,\kappa\right)  $}%
\label{m9p}%
\end{figure}

Finally, in Fig. \ref{m9p} we present phase-space portraits for the dynamical
system (\ref{ss.26}), (\ref{ss.27}) for different values of the free
parameters $\left(  \lambda,\kappa\right)  .$ Hence, it is clear that the
Minkowski universes described by points $Q_{2}$ and $Q_{4}$ are saddle points.

\section{Evolution of the perturbations}

\label{per1}

Let us now study the evolution of the cosmological perturbations around the
stationary points of special physical interests. In order to perform such
analysis we select to apply the linear perturbation theory in the Newtonian
gauge, where the perturbed spacetime around the spatially FLRW space is
written as follows \cite{amebook}
\begin{equation}
ds^{2}=a^{2}\left(  \sigma\right)  \left(  -\left(  1+2\Phi\left(
\sigma,x,y,z\right)  \right)  d\sigma^{2}+\left(  1-2\Phi\left(
\sigma,x,y,z\right)  \right)  \left(  dx^{2}+dy^{2}+dz^{2}\right)  \right)
,\label{pr.01}%
\end{equation}
where $\sigma$ is the conformal time $\sigma=\int a^{-1}dt$. Function
$\Phi\left(  \sigma,x,y,z\right)  $ includes the perturbation terms. \ As far
as the two scalar fields $\phi~$and $\psi$, are concerned, they are perturbed
as~$\phi+\delta\phi~,~\psi+\delta\psi$ where we select $\varphi\left(
\sigma,x,y,z\right)  =\delta\phi$ and $\xi\left(  \sigma,x,y,z\right)
=\delta\psi$.

In the linear perturbation theory in the Newtonian gauge the perturbative
components of Einstein's tensor are%
\begin{align}
\delta G_{0}^{0}  &  =-\frac{2}{a^{2}}\nabla^{2}\Phi+\frac{6}{a}%
\mathcal{H}\left(  \dot{\Phi}+a\mathcal{H}\Phi\right)  \,,\label{pr.02}\\
\delta G_{i}^{0}  &  =-\frac{2}{a^{2}}\nabla_{i}\left(  \dot{\Phi
}+a\mathcal{H}\Phi\right)  \,,\label{pr.03}\\
\delta G_{i}^{i}  &  =2\Phi\left(  \frac{2}{a}\mathcal{\dot{H}}+3\mathcal{H}%
^{2}\right)  +\frac{2}{a^{2}}\left(  \ddot{\Phi}+3a\mathcal{H}\dot{\Phi
}\right)  \,, \label{pr.04}%
\end{align}
where now dot means derivative with respect the variable $\sigma$, and
$\mathcal{H}=\frac{\dot{a}}{a^{2}}$ is the Hubble parameter in the new frame
and $i,j,k=1,2,3$.

The contribution of the scalar fields in the linear perturbations $\delta
T_{\mu\nu}$ is as follows%

\begin{equation}
\delta T_{0}^{0}=-\frac{1}{a^{2}}\left(  \dot{\phi}\dot{\varphi}-\dot{\phi
}^{2}\Phi\right)  -V_{,\phi}\varphi+\frac{e^{\kappa\phi}}{a^{2}}\left(
\dot{\psi}\dot{\xi}-\dot{\psi}^{2}\Phi\right)  +\frac{e^{\kappa\phi}}{2a^{2}%
}\varphi\dot{\psi}^{2} \label{pr.07}%
\end{equation}%
\begin{equation}
\delta T_{i}^{\left(  d\right)  0}=\frac{1}{a^{2}}\nabla_{i}\left(  \dot{\phi
}\varphi-e^{\kappa\phi}\dot{\psi}\xi\right)  . \label{pr.08}%
\end{equation}%
\begin{equation}
\delta T_{i}^{i}=\frac{1}{a^{2}}\left(  \dot{\phi}\dot{\varphi}-\dot{\phi}%
^{2}\Phi\right)  -V_{,\phi}\varphi-\frac{e^{\kappa\phi}}{a^{2}}\left(
\dot{\psi}\dot{\xi}-\dot{\psi}^{2}\Phi\right)  +\frac{e^{\kappa\phi}}{2a^{2}%
}\varphi\dot{\psi}^{2} \label{pr.08a}%
\end{equation}

Furthermore, from the equation of motions (\ref{ss.07}), (\ref{ss.08}), that
is, the conservation law $T_{~~~;\nu}^{\mu\nu}=0$, we find for the following
system which describes the solution of the perturbations%
\begin{equation}
\frac{1}{a^{2}}\left(  \ddot{\varphi}-\nabla^{2}\varphi-4\dot{\Phi}\dot{\phi
}\right)  +\frac{2}{a}\mathcal{H}\dot{\varphi}+\frac{\kappa^{2}}{2a^{2}%
}e^{\kappa\phi}\dot{\psi}^{2}\varphi+V_{,\phi\phi}\varphi+2\Phi V_{,\phi
}=0,\label{pr.10}%
\end{equation}%
\begin{equation}
\frac{1}{a^{2}}\left(  \ddot{\xi}-\nabla^{2}\xi-4\dot{\Phi}\dot{\psi}\right)
+\frac{2}{a}\mathcal{H}\dot{\xi}+\frac{\kappa}{a^{2}}\dot{\psi}\dot{\varphi
}=0\,,\label{pr.11}%
\end{equation}

We define the new independent variable $\tau$ such that,
\begin{align}
\frac{1}{a}\frac{d}{d\sigma}  &  =\sqrt{1+\mathcal{H}^{2}}\frac{d}{d\ln a},\\
\frac{1}{a^{2}}\frac{d^{2}}{d\sigma^{2}}  &  =\left(  1+\mathcal{H}%
^{2}\right)  \frac{d^{2}}{d\tau^{2}}+\mathcal{HH}^{\prime}%
\end{align}
where a prime means total derivative with respect to $\tau$, i.e.
$\mathcal{H}^{\prime}=\frac{d\mathcal{H}}{d\tau}$.

In the new variables, equations (\ref{pr.10}) and (\ref{pr.11}) become%
\begin{equation}
\varphi^{\prime\prime}+\eta\left(  \frac{1-3w_{eff}}{2}\right)  \varphi
^{\prime}-4\sqrt{6}\Phi^{\prime}x+\frac{k^{2}}{a^{2}\left(  1+\mathcal{H}%
^{2}\right)  }\varphi+3\kappa^{2}z^{2}\varphi+3\lambda y^{2}\left(
\lambda\Gamma\left(  \lambda\right)  \varphi+2\Phi\right)  =0\label{pr.12}%
\end{equation}%
\begin{equation}
\xi^{\prime\prime}+\eta\left(  \frac{1-3w_{eff}}{2}\right)  \xi^{\prime
}-4\sqrt{6}\Phi^{\prime}e^{-\frac{\kappa}{2}\phi}z+\frac{k^{2}}{a^{2}\left(
1+\mathcal{H}^{2}\right)  }\xi+\sqrt{6}\kappa\varphi ze^{-\frac{\kappa}{2}%
\phi}=0.\label{pr.13}%
\end{equation}
in which we replaced $\varphi=\varphi\left(  \tau\right)  e^{-ikx^{i}}$,
$\xi=\xi\left(  \tau\right)  e^{-ikx^{i}}$ and $\Phi=\Phi\left(  \tau\right)
e^{-ikx^{i}}$ and we assumed $\eta=\frac{\mathcal{H}}{\sqrt{1+\mathcal{H}^{2}%
}},$ while we have replaced~$\eta\frac{\mathcal{H}^{\prime}}{\mathcal{H}%
}=-\frac{3}{2}\left(  1+w_{eff}\right)  \,.$

In addition, from the field equations we find
\begin{equation}
\Phi^{\prime}=-\frac{\sqrt{6}}{2}\left(  x\varphi-e^{\frac{\kappa}{2}\phi}%
z\xi\right)  -\eta\Phi. \label{pr.14}%
\end{equation}%
\begin{equation}
-\frac{2k^{2}}{a^{2}\sqrt{1+\mathcal{H}^{2}}}\Phi+6\left(  \eta\Phi^{\prime
}+\eta^{2}\Phi\right)  =-\left(  \sqrt{6}x\varphi^{\prime}-6x^{2}\Phi\right)
-3\lambda y^{2}\varphi+\left(  \sqrt{6}e^{\frac{\kappa}{2}\phi}z\xi^{\prime
}-6z^{2}\Phi\right)  +3\varphi z^{2}. \label{pr.14a}%
\end{equation}

We proceed by doing the change of variables $\xi=\zeta e^{-\frac{\kappa}%
{2}\phi}$, thus equations (\ref{pr.13}), (\ref{pr.14}) and (\ref{pr.14a}) are
written as follows%
\begin{align}
0  &  =\zeta^{\prime\prime}-\frac{\sqrt{6}\kappa}{2}\zeta^{\prime}%
x+\frac{\kappa^{2}\sqrt{6}}{4}\zeta x^{\prime}+\frac{k^{2}}{a^{2}\left(
1+\mathcal{H}^{2}\right)  }\zeta-4\sqrt{6}\Phi^{\prime}z\label{pr.16}\\
&  +\left(  \eta\left(  \frac{1-3w_{eff}}{2}\right)  \right)  \left(
\zeta^{\prime}-\frac{\sqrt{6}\kappa}{2}\zeta x\right)  +\sqrt{6}\kappa\varphi
z,\nonumber
\end{align}%
\begin{equation}
\Phi^{\prime}=-\frac{\sqrt{6}}{2}\left(  x\varphi-z\zeta\right)  -\eta\Phi,
\label{pr.17}%
\end{equation}
and%
\begin{align}
0  &  =-\left(  \sqrt{6}x\varphi^{\prime}-6x^{2}\Phi\right)  -3\lambda
y^{2}\varphi+\left(  \sqrt{6}z\left(  \zeta^{\prime}-\frac{\sqrt{6}\kappa}%
{2}\zeta x\right)  -6z^{2}\Phi\right) \label{pr.18}\\
&  +3\varphi z^{2}+\frac{2k^{2}}{a^{2}\sqrt{1+\mathcal{H}^{2}}}\Phi-6\left(
\eta\Phi^{\prime}+\eta^{2}\Phi\right)  .\nonumber
\end{align}

Until now we have considered variables at the finite regime. From the results
of the previous section, we found that at the infinity the stationary points
have the same physical properties as that of the finite regime. Hence, in
order to study the evolution of the perturbations at the asymptotic solutions
we select to work with the finite variables only.

\subsection{Perturbations at the stationary points $P_{7}$ and $P_{8}$}

We carry on our analysis by studying the qualitative evolution for the
perturbations on the background space in which the physical solution is
described by the asymptotic solutions at the stationary points. We observe
that equations (\ref{pr.12}), (\ref{pr.17}) and (\ref{pr.18}) are independent
on $\zeta$ when the second field does not contribute in the background space,
that is, for $z=0$, the evolution of the perturbations it is similar with that
for the quintessence scalar field. Hence, we shall on the new stationary
points which describe the hyperbolic inflation, points $P_{7}$ and $P_{8}$.
For the exponential potential $\Gamma\left(  \lambda\right)  =1,$ while at the
stationary points $P_{7}$ and $P_{8}$ we find $\eta^{2}=1$, which means that
$\mathcal{H}^{2}\rightarrow+\infty.$

Consequently, the perturbation equations for the scalar fields read%
\begin{align}
0  & =\varphi^{\prime\prime}+\frac{(1-3w_{eff})}{2}\eta\varphi^{\prime
}+3\kappa^{2}z^{2}\varphi\nonumber\\
& +\frac{\lambda y^{2}\left(  -3\varphi\left(  (\lambda+1)z^{2}-\lambda\left(
x^{2}+y^{2}\right)  \right)  +\sqrt{6}\left(  x\varphi^{\prime}-z\zeta
^{\prime}\right)  +3\kappa xz\zeta\right)  }{x^{2}-z^{2}}+\nonumber\\
& +\frac{2\left(  \varphi\left(  6x^{3}-6xz^{2}+\sqrt{6}\eta\left(  \lambda
y^{2}-z^{2}\right)  \right)  +\zeta\left(  -6x^{2}z+\sqrt{6}\eta\kappa
xz+6z^{3}\right)  +2\eta x\varphi^{\prime}-2\eta z\zeta^{\prime}\right)
}{x^{2}-z^{2}},\label{pr.19}%
\end{align}%
\begin{align}
0  & =\zeta^{\prime\prime}+\frac{(1-3w_{eff})}{2}\eta\left(  \zeta^{\prime
}-\frac{\sqrt{6}}{2}\kappa x\zeta\right)  -\sqrt{\frac{3}{2}}\kappa
x\zeta^{\prime}+2\sqrt{6}\kappa z\varphi+\nonumber\\
& -\frac{2z\left(  \varphi\left(  6x^{3}-6xz^{2}+\sqrt{6}\eta\left(  \lambda
y^{2}-z^{2}\right)  \right)  +z\zeta\left(  -6x^{2}+\sqrt{6}\eta\kappa
x+6z^{2}\right)  +2\eta x\varphi^{\prime}-2\eta z\zeta^{\prime}\right)
}{z^{2}-x^{2}}.\label{pr.20}%
\end{align}

We write the two second-order differential equations as a linear system
\begin{align}
\varphi^{\prime}  & =p_{\varphi}~,~\label{pr.21}\\
\zeta^{\prime}  & =p_{\zeta}\label{pr.22}\\
p_{\varphi}^{\prime}  & =f\left(  \varphi,\zeta,p_{\varphi},p_{\zeta}\right)
\label{pr.23}\\
p_{\zeta}^{\prime}  & =g\left(  \varphi,\zeta,p_{\varphi},p_{\zeta}\right)
~\label{pr.24}%
\end{align}

\begin{figure}[ptb]
\centering\includegraphics[width=0.9\textwidth]{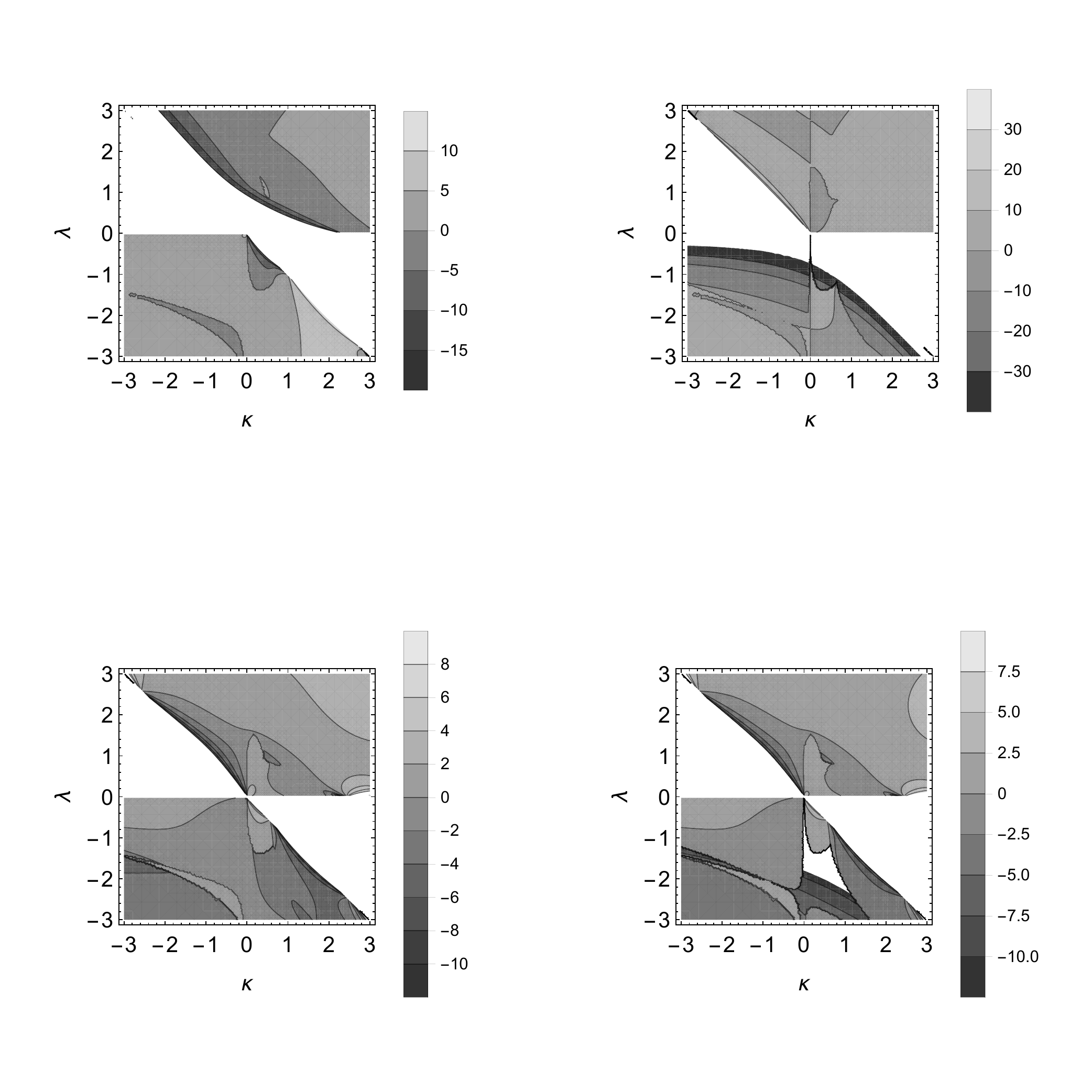} \caption{Contour
plots for the real part for the four eigenvalues of the linear system
(\ref{pr.21})-(\ref{pr.24}) in the space of the free variables $\left\{
\kappa,\lambda\right\}  $ with background solution described by $P_{7}$.}%
\label{fi10}%
\end{figure}

\begin{figure}[ptb]
\centering\includegraphics[width=0.9\textwidth]{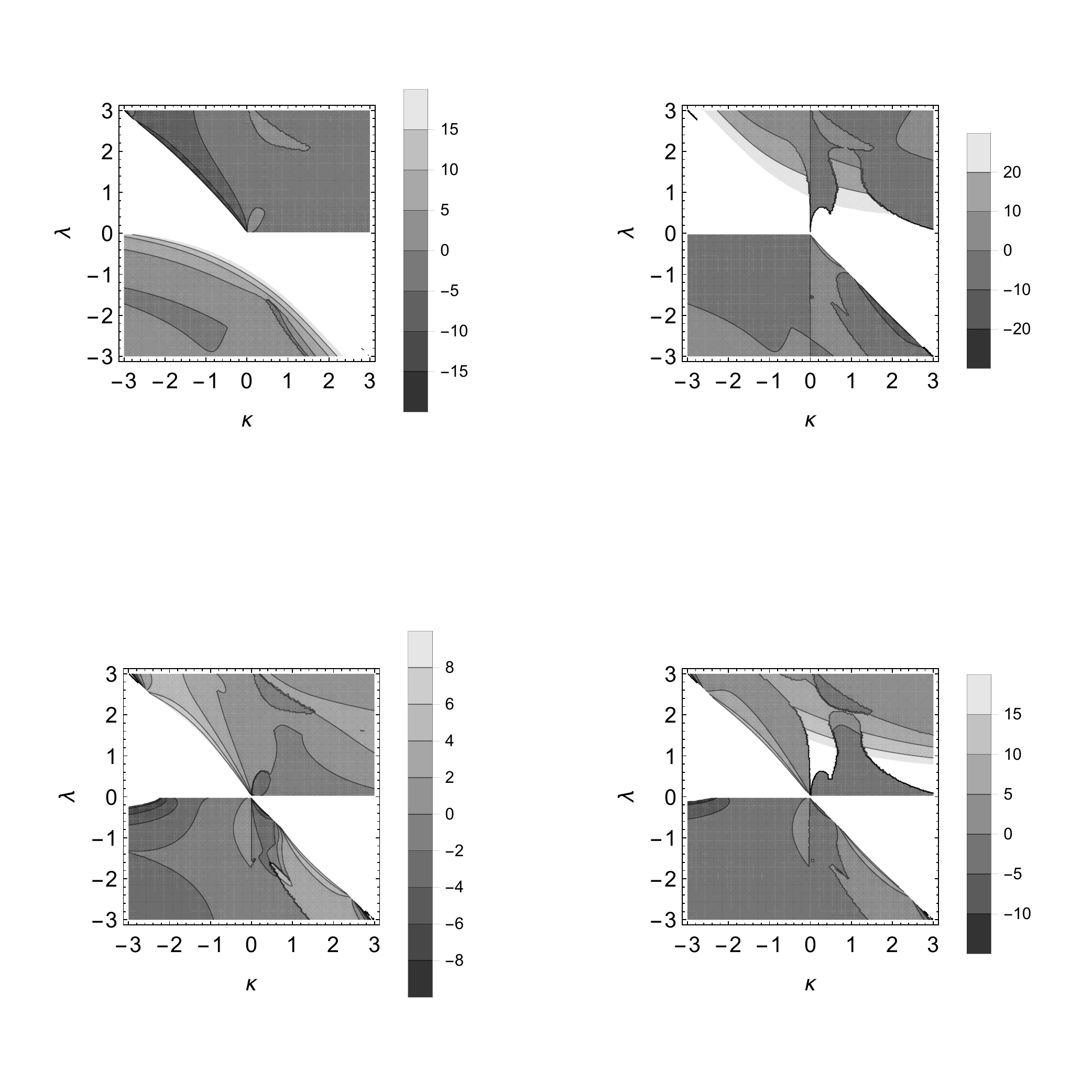} \caption{Contour
plots for the real part for the four eigenvalues of the linear system
(\ref{pr.21})-(\ref{pr.24}) in the space of the free variables $\left\{
\kappa,\lambda\right\}  $ with background solution described by $P_{8}$.}%
\label{fi11}%
\end{figure}

For the background solutions described by the stationary points $P_{7}$ and
$P_{8}$, the dynamical system (\ref{pr.21})-(\ref{pr.24}) admit the stationary
point $\left(  \varphi,\zeta,p_{\varphi},p_{\zeta}\right)  =\left(
0,0,0,0\right)  $. The real part of the eigenvalues for the linear system are
plot numerically. In Fig. \ref{fi10} we plot the real part of the eigenvalues
for the system (\ref{pr.21})-(\ref{pr.24}) at the point $P_{7}$, while in Fig.
\ref{fi11} we plot the eigenvalues for the background solution described by
$P_{8}$.

\begin{figure}[ptb]
\centering\includegraphics[width=0.9\textwidth]{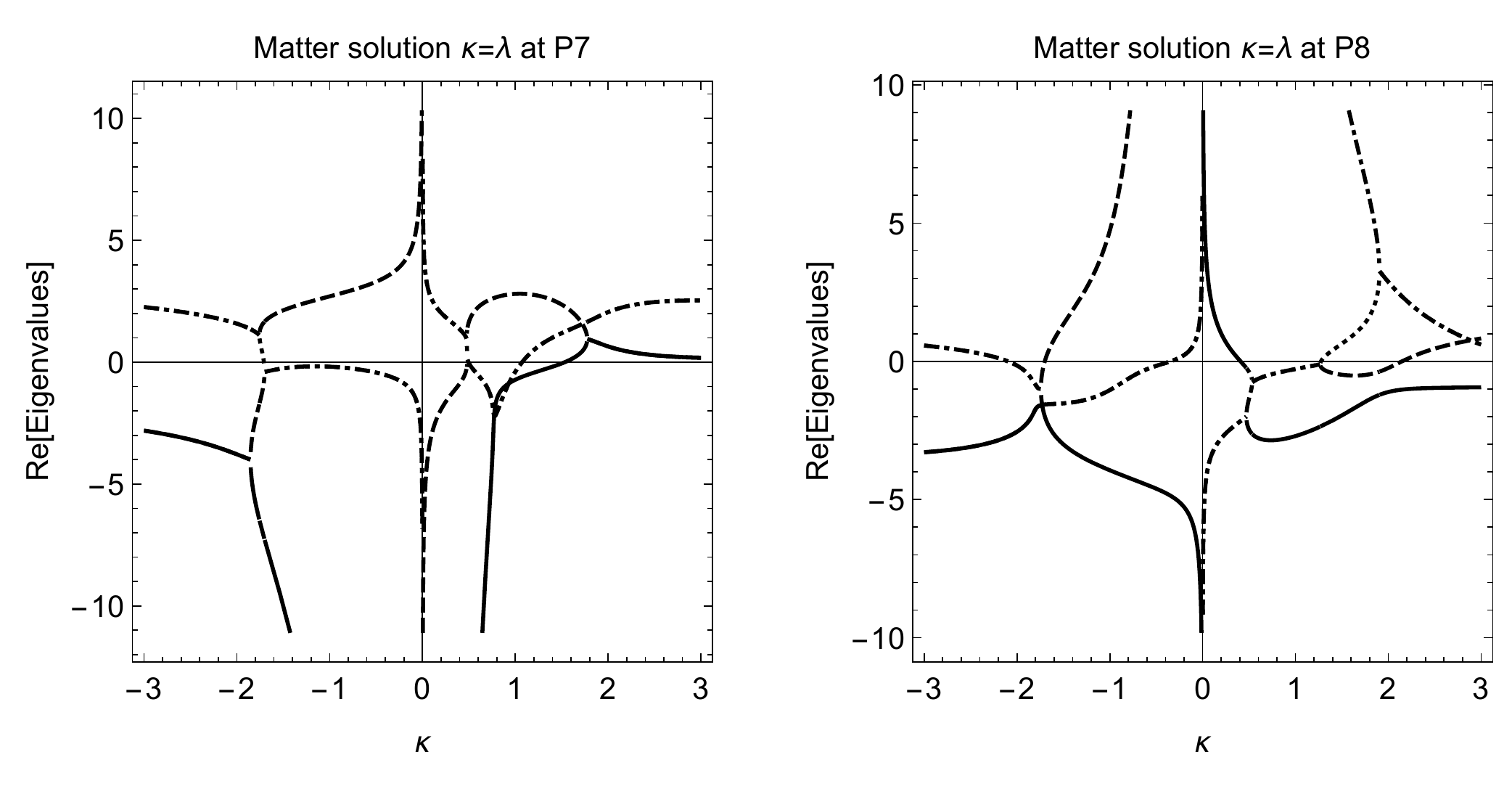} \caption{Real part
for the four eigenvalues of the linear system (\ref{pr.21})-(\ref{pr.24}) in
in the case where $P_{7}$ and $P_{8}$ describe the matter era. }%
\label{fi12}%
\end{figure}

The equation of state parameter for the background solution at the stationary
points $P_{7}$ and $P_{8}$ is $w_{eff}=1-\frac{2\kappa}{\kappa+\lambda}$. Thus
for $\kappa=\lambda$, $w_{eff}=0$ which means that the solution describes the
matter era. The real part for the eigenvalues of the perturbation system in
the matter solution with $\kappa=\lambda$ is presented in Fig. \ref{fi12}. We
observe that not all the eigenvalues are negative which means that the
perturbations do not decay.

\section{Conclusions}

\label{con00}

In this work, we have investigated the global dynamics of a multiscalar field
cosmological model known as hyperbolic Chiral-Phantom theory. It is a
two-scalar field theory in which the scalar fields are minimally coupled to
gravity, but they interact in the kinetic part. In particular, the kinetic
Lagrangian for the scalar fields defines a two-dimensional manifold of
constant negative curvature, i.e., a hyperbolic plane. Moreover, one of the
two scalar fields is assumed to have a negative kinetic term, which means that
its energy density can be negative and have phantom properties. This model has
been recently proposed in \cite{and3} and it was found that provides a
different cosmological history from the standard hyperbolic Chiral theory.
While there are similarities with the quintom theory, the two theories are dissimilar.

To study the cosmological history provided by this model, we use new
dimensionless variables different from the usual $H$-normalization because the
Hubble function can change the sign and vanishes, so, the $H$-normalization is
not applicable. In terms of the new variables, the field equations were written
in the equivalent form of an algebraic-differential system. For this system, we
have investigated the stationary points and their stability. Each stationary
point describes a specific epoch in the cosmological evolution. The stationary
points were investigated in the local variables, also at the infinity region
with the use of Poincar\'{e} variables. Furthermore, for the scalar field
potential, we have considered the exponential potential.

In the finite regime with the use of local variables, we found that the
dynamical system admits nine stationary points. That is a greater number from
that of the quintessence model, and of the standard hyperbolic model. Points
$P_{1},~P_{2}$,~$P_{3}$,~$P_{4}$ and $P_{5},~P_{6}$ can be categorized as
$\left\{  P_{1},P_{3}\right\}  ,~\left\{  P_{2},P_{4}\right\}  ~$and $\left\{
P_{5},P_{6}\right\}  $. These three categories describe the three stationary
points of the quintessence scalar field model with exponential potential
\cite{cop1}. The two first categories describe the stiff fluid solutions,
while the third category describes the scaling solution. Moreover, from the
stability analysis of the stationary points, we found that the quintessence
model is possible to be an attractor for the system. Stationary points
$P_{7},~P_{8}$ describe the so-called hyperbolic inflation, while $P_{9}$
corresponds to Einstein's universe in the spatially flat FLRW background
space, that is, the Minkowski space. Point $P_{9}$ was found to be a saddle
point. $P_{7}$ can be an attractor while $P_{8}$ is a saddle point.

However, by using the Poincar\'{e} variables we were able to find new
stationary points. Specifically, we found ten families of stationary points.
Points $\left\{  Q_{1}^{\pm},Q_{3}^{\pm},Q_{5}^{\pm},Q_{6}^{\pm}\right\}  $
describe the limit of quintessence theory, point $\left\{  Q_{2}%
,Q_{4}\right\}  $ describe the Einstein static space; while the stationary
points $\left\{  Q_{7}^{\pm},Q_{8}^{\pm},Q_{9}^{\pm},Q_{10}^{\pm}\right\}  $
describe the hyperbolic inflation similarly to the points $P_{7}$ and $P_{8}$.

Furthermore, we investigated the linear cosmological perturbations in the
Newtonian gauge. We derived the perturbed field equations and the equations of
motions for the evolution of the perturbation terms for the scalar fields. The
stability properties for the scalar field perturbations were investigated at the
background solutions where the two scalar fields contribute to the
cosmological fluid.

The theory provides a rich cosmological history, either in
the finite or the infinity regime. That is an important property because there
is more than one set of initial conditions which can be the same cosmological
evolution. Furthermore, this model can play a role not only
in the description of inflation, as it is mainly used, but also can be a dark
energy candidate. In future work, we plan to investigate further considerations
by using cosmological observations.

\begin{acknowledgments}
The research of AP and GL was funded by Agencia Nacional de Investigaci\'{o}n
y Desarrollo - ANID through the program FONDECYT Iniciaci\'{o}n grant no.
11180126. Additionally, GL was funded by Vicerrector\'{\i}a de
Investigaci\'{o}n y Desarrollo Tecnol\'{o}gico at Universidad Catolica del
Norte. This work is based on the research supported in part by the National
Research Foundation of South Africa (Grant Numbers 131604). Ellen de Los
Milagros Fern\'andez Flores is acknowledged for proofreading this manuscript
and for improving the English.
\end{acknowledgments}

\end{document}